\documentstyle[preprint,aps,prl]{revtex}       

\begin{document}
\tightenlines
\draft
\title{ Quantized Orbits and Resonant Transport} 
\author{Indubala I. Satija\footnote{email:isatija@sitar.gmu.edu}}
\address{
 Department of Physics, George Mason University,
 Fairfax, VA 22030}
\author{ Bala Sundaram\footnote{email:bas@math.csi.cuny.edu} }
\address{ Department of Mathematics,
CSI-CUNY,
2800, Victory Boulevard
Staten Island, NY 10314}
\date{\today}
\maketitle
\begin{abstract}
A tight binding representation of the kicked 
Harper model is used to obtain an integrable
semiclassical Hamiltonian consisting of degenerate "quantized" orbits. 
New orbits appear when renormalized Harper
parameters cross integer multiples of $\pi/2$.
Commensurability relations between the orbit frequencies
are shown to correlate with the emergence of accelerator
modes in the classical phase space of the original kicked problem. The
signature of this resonant transport is seen
in both classical and quantum behavior.
An important feature of our analysis is the emergence of a natural scaling
relating classical and quantum couplings which is necessary for
establishing correspondence.

\end{abstract}

\pacs{03.65.Sq,05.45.Mt,64.60.Ak}
\narrowtext

The search for evidence of complexity in quantum systems
when the corresponding classical system exhibits chaotic dynamics is an
active frontier in nonlinear science. Kicked systems with classically chaotic
dynamics have provided useful theoretical~\cite{Casati,Fishman} and 
experimental~\cite{expt} paradigms for elucidating the fundamental
issues of quantum chaos. Systems where quantum states can be either
extended or localized provide the added challenge of finding
signatures of the localization-delocalization transition in the
corresponding classical system. In this letter, we consider the kicked
Harper system to establish precisely such a correspondence, within the context of
fascinating issues such as anomalous transport~\cite{GZ} and 
dynamical localization~\cite{Fishman}.

The traditional approach starts with the classical analysis and then
searches for signatures in the quantum behavior. By contrast, we begin
with the quantum model and, using a novel semiclassical approach,
predict a series of enhancements to transport in the
system. These `phase transitions' are shown to correspond to ballistic
to diffusive transitions in both classical and
quantum dynamics, and are due to classical accelerator
modes. These modes, in turn, are known to result in anomalous transport~\cite{GZ}.
Our methodology is applicable to a variety of kicked systems. However, we will
illustrate our method in the kicked Harper model, 
\begin{equation}
H=L\cos(p)+K \cos(q) \sum_{k} \delta(t-k) \; ,
\label{KHamil}
\end{equation}
where the relative values of the parameters $K$ and $L$ determine the dynamics.
The classical dynamics reduces to the two-dimensional area preserving Harper map,
\begin{eqnarray}
p_{k+1}&=&p_k + K \sin(q_k)\nonumber \\
q_{k+1} & = & q_k - L \sin(p_{k+1})\;.
\label{KHMap}
\end{eqnarray}
This mapping is analogous to the
standard map or kicked rotor, with the dynamics becoming globally chaotic when both
$K$ and $L$ cross critical values. However, neither
system is ergodic as islands, arising via bifurcations, exist in the
chaotic sea leading to the possibility of anomalous transport. In particular,
hierarchical island chains for $K=L$ in the vicinity
of $2\pi$ have been shown to result in fractional
kinetics~\cite{GZ}. 

By contrast, the quantum dynamics of the kicked Harper model is quite
different from that of the quantum kicked  rotor. Most significant is the 
existence of both extended and localized states depending on $K$ and
$L$. Earlier studies indicate that extended states exist for $K>L$
while localization occurs for $K<L$~\cite{Lima}, even though the
$K=L$ critical line, demarcating the onset of localization, has been 
rigorously established only in the absence of kicking.
The existence of this transition allows the possibility of both
ballistic as well as diffusive transport in the system.

Our analysis emphasizes the dynamics along the line $K=L$ and,
using a novel semiclassical approach, predicts a rich variety of
behavior. The key is the mapping of the
temporal dynamics of the quantum kicked model to a 
continuum time-independent integrable Hamiltonian, within which 
the complexity of the original problem manifests itself in terms of
degenerate classical orbits. These orbits are
indexed by an integer $n$, where the $n^{th}$ orbit appears when the 
renormalized coupling $\bar{K}=K/(2\hbar)=\bar{L}$ exceeds $n \pi/2$. 
The phase space of this semiclassical Hamiltonian resembles 
the {\it unkicked} Harper model except 
that only certain orbits are allowed. Therefore, in 
this semiclassical formulation, kicking the system {\it selects
a discrete number of classical trajectories of the
unkicked Harper system}. These
additional orbits are shown to enhance both classical and quantum
transport {\it in the kicked Harper model}. Finally,
classical and quantum results display good correspondence provided the
classical parameter is set to be  $\bar{K}$. This implies 
an $\hbar$ scaling considerably different from that used earlier~\cite{Dima}.

We start from the association of the time-dependent quantum
Hamiltonian with a tight-binding lattice problem~\cite{Fishman}. 
This is obtained by projecting individual quasienergy states,
associated with the single-kick evolution operator, onto
angular momentum eigenstates $|m>$ with eigenvalues $p=m\hbar$. The
projection coefficients $u_m$ satisfy
\begin{equation}
\sum_r J_r(\bar{K})
\sin{[\frac{1}{2}\left(\omega-2\bar{L}\cos(m\hbar)\right)-\frac{\pi r}{2}]} 
u_{m+r} = 0,
\end{equation}
where $\omega$ is the quasienergy corresponding to the eigenstate.
Following Wilkinson~\cite{Wilk}, we consider the continuum limit of
the lattice model by defining $x_m=m \hbar$ and $\psi(x_m)=\psi_m$.
Thus, $(\psi_{m+r}+\psi_{m-r})$ is replaced by $2 \cos(r \pi p) \psi(x_m)$
as $p$ is the generator of space translation.
Taking $\omega=0$, the tight-binding model can be written as
$H\psi(x)=0$, where the Hamiltonian operator $H$ is,
\begin{eqnarray}
&H&=\sin(\bar{L}\cos(x))\left[J_0(\bar{K})+2\sum (-1)^r J_{2r}(\bar{K}) \cos(2r p)\right] \nonumber \\
&+&\cos(\bar{L}\cos(x)) \left[2\sum (-1)^{r-1} J_{2r-1}(\bar{K}) \cos( (2r-1)p)\right]\; .
\end{eqnarray}
Note that the lattice problem $H\psi(x)=0$ can be viewed as
an eigenvalue problem restricted to a Hilbert
subspace defined by eigenvalue $E=0$. The terms in the square brackets
$H$ can be summed exactly to give
$\cos{(\bar{K}cos(p))}$ and $\sin{(\bar{K}cos(p))}$ respectively.
Therefore, the resulting semiclassical Hamiltonian simplifies to
\begin{equation}
H_{scl} = sin [ \bar{K} \cos(p) + \bar{L} \cos(x) ] \;.
\end{equation}

The fact that the Hamiltonian dynamics is constrained to $E=0$ implies that
the classical phase space trajectories satisfy
\begin{equation}
\bar{K} \cos(p)+\bar{L} \cos(x)=\pm n \pi,
\end{equation}
where $n$ is an integer.
This can be viewed as the ``quantization condition''
for the phase space of the unkicked Harper equation given by
$\bar{K} cos(p)+\bar{L} cos(x)=E$. Thus, within this semiclassical
formulation, the kicking effectively quantizes the classical orbits.
Another interesting feature, which we exploit later, is the fact 
that $\hbar$ appears only in $\bar{K}$ and $\bar{L}$ and not as an
independent parameter. Changing $\hbar$ simply
renormalizes the couplings $K$ and $L$. Note also that $H_{scl}$ is symmetric 
in $x$ and $p$.

The phase portrait resulting from $H_{scl}$ consists of degenerate
orbits indexed by the integer $n$. 
For $\bar{K} < \pi/2$ and
$\bar{L} < \pi/2$, there exists only a single orbit
and additional
degenerate branches appear as $\bar{K}$ and $\bar{L}$ exceed threshold values
$\bar{K}^*_n=n\pi/2$.
The quantized orbits
are {\it not necessarily closed}, unless $\bar{K}=\bar{L}$.
For the rest of the discussion, we will confine ourselves to this
symmetric case.
In the corresponding quantum problem, the condition
$\bar{K}=\bar{L}$ for closed orbits suggests a phase boundary for
a localization-delocalization transition~\cite{Wilk}.
For $\bar{K} < \pi/2$ this boundary is the separatrix labeled by $n=0$.
Therefore, $\bar{K} = \pi/2$, when the
next orbit appears, can be viewed as the semiclassical threshold for 
the onset of diffusion in the kicked Harper rotor\cite{ftresonance}. 

The frequencies of the degenerate orbits are obtained from
the equations of motion resulting from $H_{scl}$
\begin{eqnarray}
\dot{p}&=& \bar{K} (-1)^n \sin(x)\nonumber \\
\dot{x}&=& -\bar{K}(-1)^n \sin(p)\; .
\end{eqnarray}
For any $\bar{K}$, the frequency of the $n^{th}$ orbit is given, in terms of the
complete elliptic function $\cal{K}$, by 
\begin{equation}
f_n = \bar{K}/[4 \mbox{$\cal{K}$} (\alpha _{n})] \; ,
\end{equation}
where the modulus $\alpha_n= \sqrt{[1-(\bar{K}^*_n/\bar{K})^2]}$.
The turning points of these quantized orbits (for $n>0$)
$x=2\cos^{-1}(\sqrt{\bar{K}_n^*/\bar{K}})$ imply that,
at $\bar{K}=\bar{K}_n^*$, the new 
orbits appear at the origin (similarly, for $n<0$ they occur at $(\pi,\pi)$).
At threshold, the $n^{th}$ orbit frequency
$f^*_n= n/4$ is commensurate with the 
kicking frequency which was taken to be unity (see (\ref{KHamil})). 
As $\bar{K}$ increases, the frequencies increase monotonically and are, in general,
incommensurate except at special parameter values.
For example, for $\bar{K} \approx 1.4\pi$ , we have $f_1/f_2=4/3$. $f_1$ and $f_2$
are commensurate with the kicking frequency at several values:
$f_1=1/3$ at $\bar{K} \approx .86\pi$, and $f_1=1/2$ at $\bar{K} \approx 1.68\pi$ while
$f_2=2/3$ at $\bar{K} \approx 1.71 \pi$. The significance of these
`critical' $\bar{K}$ values will be seen shortly.

In general, the quantum transport is expected to be enhanced by 
tunneling between these classically distinct orbits~\cite{Wilk}.  Whenever a
threshold value $\bar{K}^*_n$ is crossed, a new branch appears which may
enhance the tunneling probability. We further expect that 
the transport may also be influenced by
the commensurability relations between the competing orbital and kicking frequencies.

Motivated by the possibility of phase transitions leading to the enhancement of
diffusion, we now investigate classical
as well as quantum transport in the original kicked Harper model.
Some of the threshold values 
predicted by our semiclassical analysis are curiously reminiscent of  known
special values $K=L=2\pi m$ beyond which
accelerator modes appear in the Harper map. These are
known to result in either ballistic or anomalous transport.
On identifying $\bar{K}$
with the classical kicking parameter, the $\bar{K}^*_n$ values include
the thresholds for accelerator modes. This scaling emerges 
naturally from our semiclassical analysis and motivates the
investigation of classical transport in the neighborhood of $K=\bar{K}_n^*=n\pi/2$.

Interestingly, a number of secondary fixed points of the classical map
(eq.(\ref{KHMap})) appear at $\bar{K}^*_n$. Of these, a large number are
accelerator modes, including the low-period ones associated with
fractional kinetics~\cite{GZ}. For example, at $\bar{K} = 2 n\pi$, four
period-1 fixed points at $(\pm \pi/2, \pm
\pi/2)$ appear on $p=q$ and $p=-q$ symmetry lines. 
Increasing $\bar{K}$ leads to pairs of period-1 points straddling the
symmetry lines and, finally, hierarchies of island chains which result
in anomalous transport. The parametric window for this behavior
extends till $\bar{K} \approx 6.6$. 
For odd-multiples of $\pi$ ($\bar{K}=(2n+1)\pi$), analogous behavior with
period-2 accelerator modes, one of which is 
$(0,\pi/2) \rightarrow (\pi/2,0) \rightarrow (0, \pi/2+2 \pi
l_K)$, is seen. The
island hierarchies at larger $K$ are exemplified in Figure.~\ref{figphase}(b).
Topological structures consisting of boundary layer island chains
and, closer to the thresholds, the more
exotic blinking and shrinking/blowing islands~\cite{GZ,GZ2} strongly suggest
anomalous transport within parametric windows.
Additional accelerator modes also appear at certain 
critical values, in particular, we have period-2
modes at $\bar{K} \approx .87 \pi$ and $\bar{K} \approx 1.38\pi$ on the $p=q$
symmetry line. Note that these are close to parameters for the commensurability 
conditions  $f_1=1/3$ and $f_1/f_2=4/3$, respectively. 

The classical behavior at $\bar{K}=(n+1/2)\pi$ is more interesting with
period-6 fixed points appearing on the symmetry lines $p=0,q=0,p=q,$ and $p=-q$. 
As illustrated by the iterates
 $(\pi/2,\pi/2) \rightarrow (\pi/2,(n+1)\pi) \rightarrow
 ((n+1)\pi,(2n+3/2)\pi) 
\rightarrow ((2n+3/2)\pi, (2n+3/2)\pi)
 \rightarrow ((2n+3/2)\pi, (n+1)\pi) \rightarrow ((n+1)\pi, \pi/2) \rightarrow
 (\pi/2, \pi/2) $, these fixed points are non-accelerating despite
 leaving the fundamental interval $[0,2\pi]$ at intermediate steps.
As $\bar{K}$ increases, these orbits also move off the symmetry lines as stable
islands, as seen from Fig.~\ref{figphase}(a). Even though the
fixed point is non-accelerating, {\it an initial condition in the vicinity of the fixed 
point returns outside the fundamental domain} after 6 iterations. In
this sense, accelerated transport can occur at these $\bar{K}$. 

The effect of these accelerator modes on classical transport 
is illustrated in Fig.~\ref{cltrans}(a) in terms of the exponent $\mu$
defined by $<p^2> \propto t^\mu$, where the braces denote an ensemble
average~\cite{mufoot}. Many spikes in $\mu$ are seen
as the parameter $\bar{K}$ is changed.
The sharp, resonant, structures corresponding to ballistic transport, $\mu 
\approx 2$, are due to the narrow ranges of stability of the
accelerator modes. Interestingly, these peaks appear very close
to the $K^*_n$ values as well as $\bar{K}$ values where the
orbital frequencies exhibit commensurability. A few of these special 
values are indicated by the arrows in the figure. Thus our quantum analysis is providing
information about the classical transport properties. Note that
for smaller $\bar{K}/\pi \approx 0.6$, a
stochastic web dominates the classical transport.

As shown in earlier work~\cite{GZ,BZ}, enhancement
of the quantum localization length $\xi$ can be used as a signature of classical 
ballistic behavior. Here we utilize this idea while, at the
same time, contrasting the scaling suggested by our analysis with that
used earlier~\cite{GZ,Dima,BZ}. The resonant transport in classical
and quantum behavior shows good correspondence provided the classical
stochasticity parameter is taken
to be $\bar{K}=K/(2\hbar)$. This is quite
different from $K_q=K sin(\hbar/2)/(\hbar/2)$ obtained
earlier~\cite{Dima} by contrasting correlations in 
the classical and quantum dynamics. $\bar{K}$ emerges naturally from a
`thermodynamic' formalism, as the tight-binding (eigenvalue) problem deals with only a
single quasienergy state~\cite{scalfoot}. As such, this may be valid beyond a certain
time scale. We consider moderate values of $\hbar$ and, as seen from
panels Fig.~\ref{cltrans}(b) and(c), spikes in $\xi$ appear close to
classical values only when $\bar{K}$, and \underline{not} $K_q$, is
considered. At these values of $\hbar$, the issue  of reaching large
$t$ behavior does not arise. Also, as in earlier work~\cite{BZ}, 
the quantization scale is much larger than the size of the classical 
structures causing the accelerated transport. The intriguing
possibility, for smaller $\hbar$, of a 
cross-over from $K_q$ to $\bar{K}$ at longer times warrants further
exploration. 

Figure~\ref{qlinesh} shows the quantum
lineshape at one of the peaks (a) contrasted with the shape away from
the peak (b). It is clear that the quantum transport at the peak
has a ballistic component which results in the well-defined
shoulders in Fig.~\ref{qlinesh}(a) as compared with panel (b).

Despite the unphysical nature of the Harper system,
resonant transport at $K^*_n$ and other critical values
can be tested experimentally. As
the origin of resonances has been related to accelerator modes, analogous behavior
exists in the kicked rotor (standard map)~\cite{expt,footnotestd} and oscillator (web
map), both of which are experimentally realizable in atom optics.
In fact, even the question of scaling as discussed above could
also be resolved experimentally in this context.

The research of IIS is supported by a grant from National Science
Foundation DMR~097535. The work of BS was supported by the National Science
Foundation PHY~9800966 and a grant from the City University of New York
PSC-CUNY Research Award Program.

\begin{figure}
\caption{Phase space portraits of the accelerator islands beyond the thresholds
  $K=\bar{K}_n^*=n\pi/2$. (a) $K=5.04$($n=3$) where one of the
  six islands is shown; (b) $K=3.199$ ($n=4$) where only
  one of four symmetric structures is shown. Note the size of these
  structures in relation to the $\hbar$ values considered.}
\label{figphase}
\end{figure}

\begin{figure}
\caption{(a)Variation of the classical transport exponent $\mu$ with
  the classical stochasticity parameter set to
  $\bar{K}$. An ensemble of $1001$ initial
  conditions along the line $p=0$ was iterated $10^5$ times, and the
  last $5\times10^4$ iterations were used to obtain $\mu$. The arrows
  at the top indicate some of the special values arising 
  from commensurability and thresholds $\bar{K}_n^*$ (see text).
 (b)-(c) Variation of the localization length $\xi$ with $\bar{K}$
 (solid) and $K_q$ (dashed). $\hbar=2\pi\sigma^M$ and (b) $M=1$ and
 (c)$M=3$, where $\sigma$ refers to the golden mean. A fixed interaction 
  time of $4000$ kicks was considered. }
\label{cltrans}
\end{figure}

\begin{figure}
\caption{Quantum lineshape at the end of $4000$ kicks. (a)
  $\bar{K}=6.4402648$ while (b)
  $\bar{K}=6.3146012$. $\hbar=2\pi\sigma^3$ and case (a) is at a peak in
  Figure 2(c) while (b) is off the peak.}
\label{qlinesh}
\end{figure}


\begin{references}
\bibitem{ftresonance} This value is reminiscent of the threshold value
for the onset of global diffusion obtained from the resonance overlap criterion.
\bibitem{mufoot} $\mu$ simply reflects the existence of ballistic
  transport over some fraction of the time period considered, which is
  sufficient for our purposes.
\bibitem{scalfoot} With this view, the singularity as $\hbar \rightarrow
0$ may reflect the well-known non-commutativity of this limit with the  $t \rightarrow
\infty$ limit.
\bibitem{footnotestd} Many of the semiclassical
predictions discussed here are also valid for standard kicked
rotor. The thresholds change to integer
multiples of $\pi$ where an open trajectory becomes closed, proving a new
competing frequency in the semiclassical Hamiltonian.
\end{references}
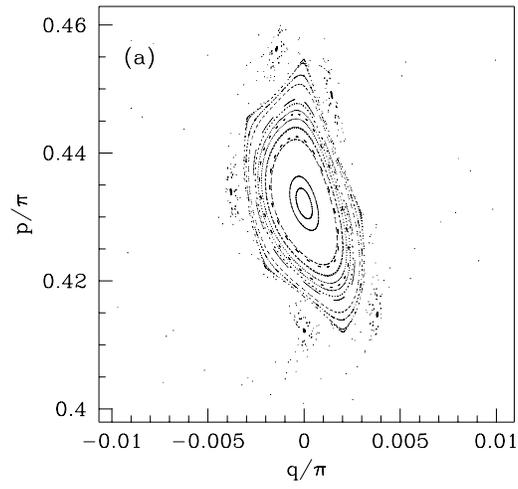
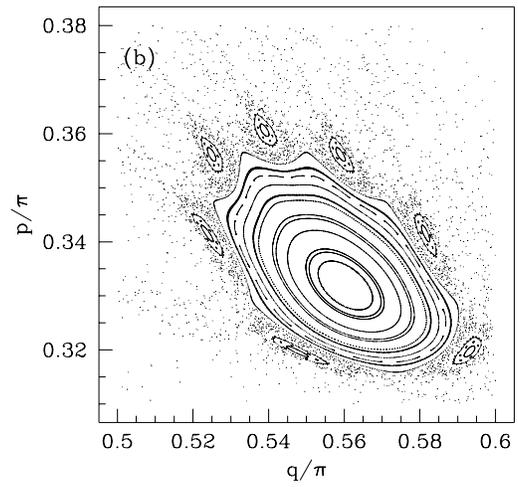
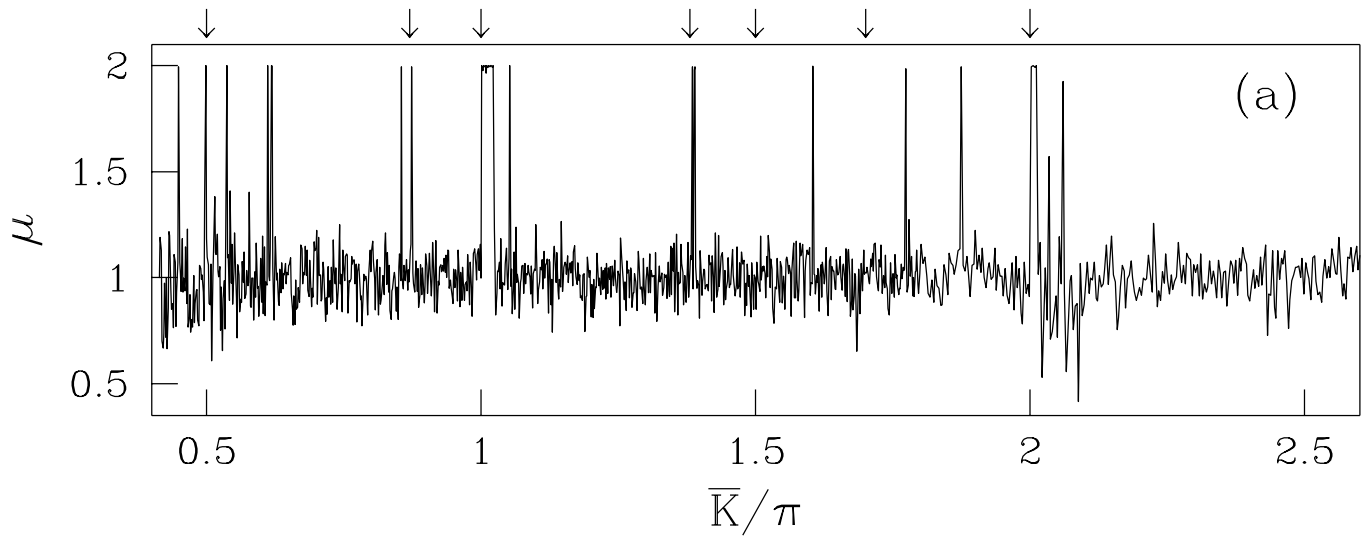
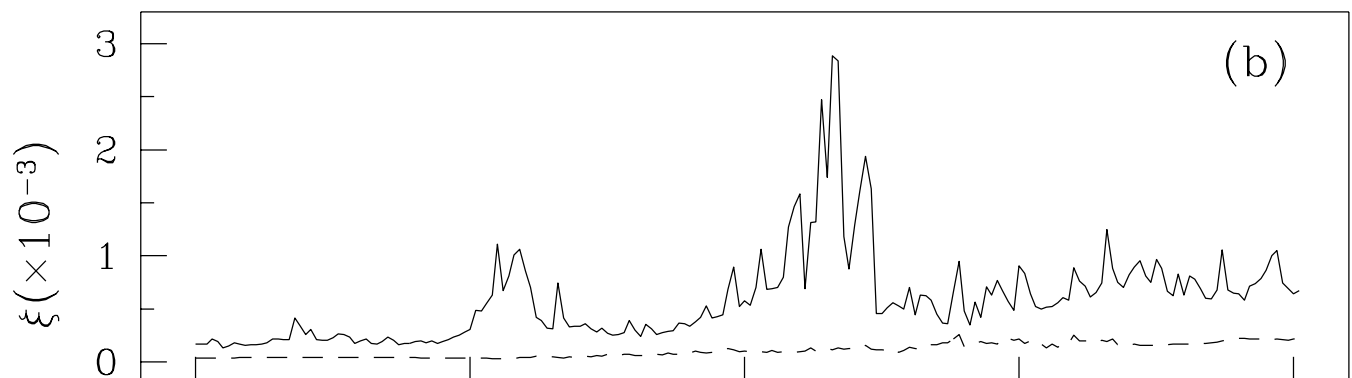
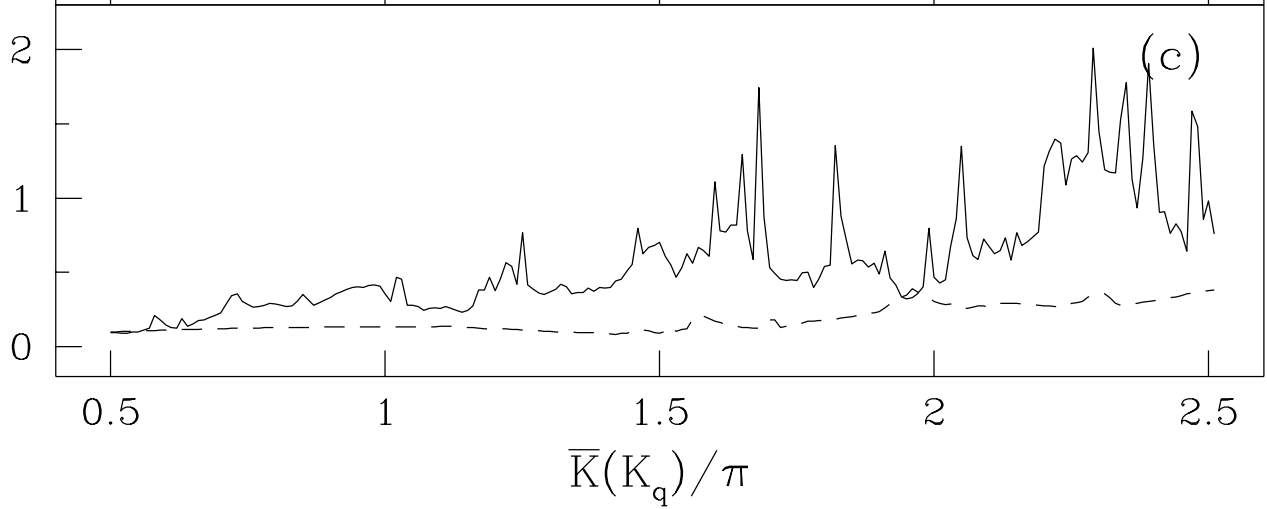
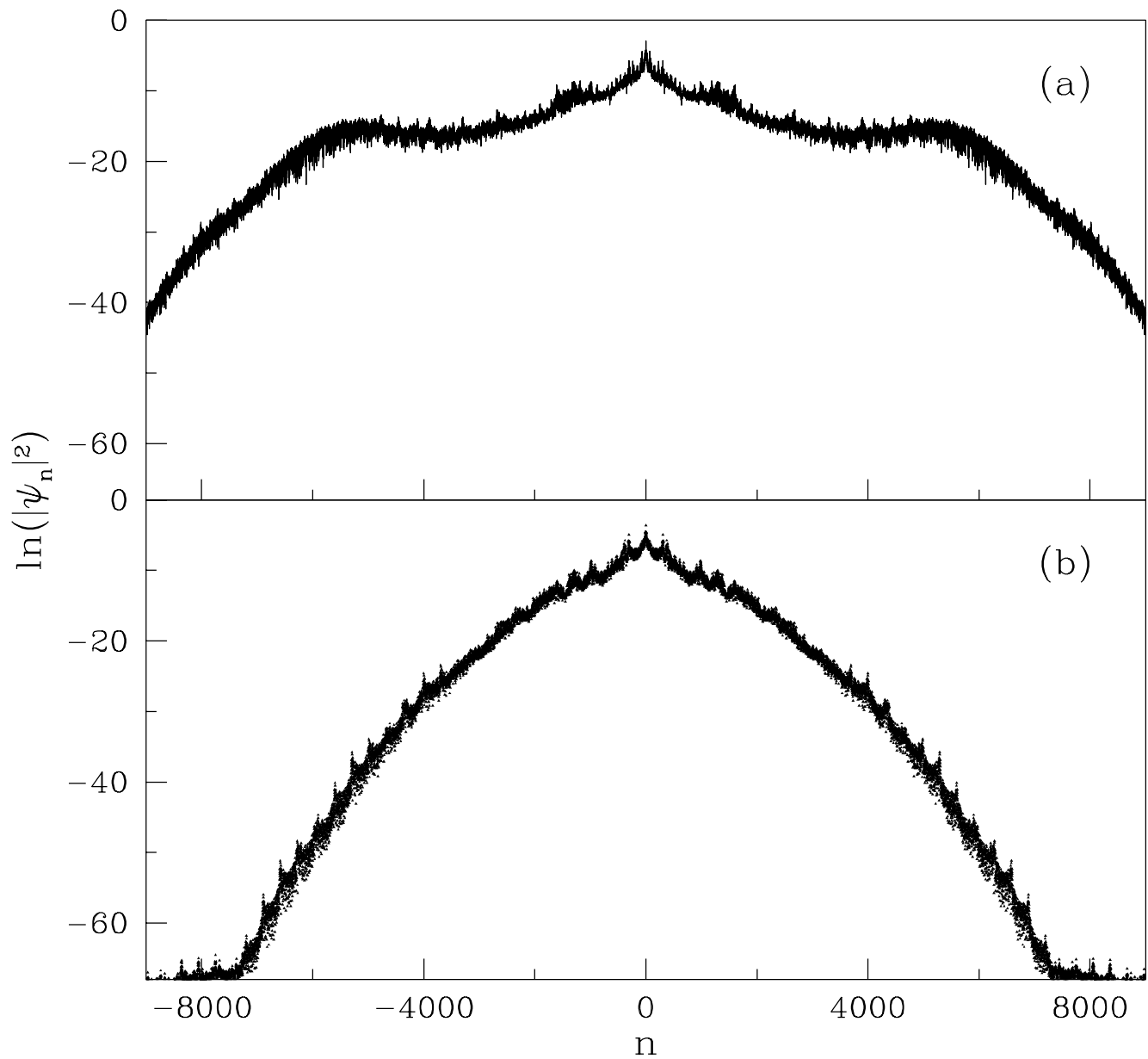
\end{document}